\documentclass{article}
\usepackage[preprint]{neurips_2026}
\usepackage[utf8]{inputenc}
\usepackage[T1]{fontenc}
\usepackage{amsmath,amssymb,booktabs,graphicx,microtype,array}
\usepackage{xcolor,xurl,needspace}
\usepackage[hidelinks]{hyperref}
\newcommand{\code}[1]{\texttt{#1}}
\title{Jevolution: A Framework for Studying\\Communication and Collective Dynamics\\in AI Agents}
\ifdefined\anonymousdraft
\author{Anonymous authors}
\else
\author{\begin{tabular}{@{}ccc@{}}
Xirui Huang & Nikhil Hooda & Utkarsh Gupta\\
\normalfont University of Waterloo & \normalfont University of Waterloo & \normalfont University of Waterloo\\
\normalfont\small\href{mailto:x36huang@uwaterloo.ca}{\texttt{x36huang@uwaterloo.ca}} &
\normalfont\small\href{mailto:n7hooda@uwaterloo.ca}{\texttt{n7hooda@uwaterloo.ca}} &
\normalfont\small\href{mailto:u2gupta@uwaterloo.ca}{\texttt{u2gupta@uwaterloo.ca}}
\end{tabular}}
\fi
\begin{document}
\maketitle
\begin{center}\small Preprint. Revised 1 October 2026.\\
\ifdefined\anonymousdraft
Project links withheld in this anonymous manuscript.
\else
\href{https://jevolution.world}{Project website}\quad\textbar\quad
\href{https://github.com/utkarshg20-org/jevolution}{Source repository}
\fi
\end{center}
\begin{abstract}
Experiments with remote decision-making agents depend on more than the policy: the world may change during inference, the simulator determines which actions are available, and births and immigration alter who can act. We present Jevolution, an instrumented spatial predator--prey framework that makes these dependencies explicit. Its reference implementation separates the evolving world, asynchronous request scheduling, controller adapters, and experiment records. The model combines local observations, energy and resource constraints, predefined signals, and inherited scalar traits. We describe these mechanisms using the ODD protocol and distinguish implemented controls from research questions requiring additional interventions. Free synthetic fixtures exercise scheduling, signaling, and inheritance against the pinned implementation; they validate specified software behavior rather than model intelligence or ecological realism. A completed communication case study illustrates the end-to-end workflow using Jev to control both species. 
Across 6 complete habitat-seed pairs, the mean signals-on minus signals-off difference in normalized rabbit abundance was -0.0095 (nominal 95\% pair-bootstrap interval [-0.0772, 0.0602]). This small study does not resolve the direction of the signal-system effect.
 The contribution is an inspectable experimental system and its evidence boundaries, not a new policy-learning algorithm or a demonstration of emergent language, adaptive selection, or biological predictive validity.
\end{abstract}

\section{Introduction}
A simulation can change while an agent is deciding what to do. A resource may disappear, a target may move, and a warning may expire before a response arrives. In a population of agents, the rate at which decisions become available also depends on who is alive, who is eligible to act, and how requests share finite concurrency. These details can change the meaning of a policy comparison. Real-time and delayed reinforcement learning provide precedents for treating timing as part of the interaction process \citep{ramstedt2019realtime,ramstedt2020delays}.

Jevolution provides an artificial environment in which these dependencies can be inspected together. Rabbits and wolves act on a spatial habitat with food, water, terrain, predation, reproduction, and immigration. A controller chooses from engine-generated actions while the world continues to advance. Rabbit signals and inherited traits provide mechanisms for posing questions about communication and population change. They do not, merely by existing, establish emergent coordination or natural selection.

Our central systems question is: \emph{what must be exposed and recorded to make experiments with asynchronous, interacting populations interpretable?} We answer through a concrete implementation, explicit process semantics, executable engineering checks, and a completed case study. The reference task is predator--prey; we do not claim a general simulator for arbitrary social or biological systems. The intended user is a researcher who needs to inspect observation construction, schedule policy calls, modify a mechanism, and trace a population outcome back to a reproducible run record.

The paper contributes: (1) an architecture and ODD specification connecting world dynamics, controller choices, scheduling, and records (Sections~\ref{sec:architecture}--\ref{sec:model}); (2) an evidence map for timing, communication, emergence, and inheritance, separating implemented mechanisms from unresolved hypotheses (Section~\ref{sec:questions}); and (3) software-level validation plus a paired communication case study demonstrating the experimental workflow (Section~\ref{sec:evaluation}). The source and analysis are pinned locally. Figure~\ref{fig:interface} shows the interactive interface; Appendix~\ref{app:visual-overview} illustrates recorded experimental states. The contribution is this concrete integration and its inspectability; no superiority over existing platforms, new learning algorithm, or independently adopted benchmark is established.

\section{Related work}\label{sec:related}
\paragraph{LLM agents and simulation.}
Generative Agents combines memory, reflection, and planning in an interactive social world \citep{park2023generative}. Project Sid describes a real-time architecture for populations of agents in Minecraft \citep{altera2024sid}, while AgentSociety integrates agents, an environment, and a simulation engine for social interventions \citep{piao2025agentsociety}. Affordable Generative Agents addresses the cost of repeated inference through reusable behavior and compressed context \citep{yu2024affordable}. A survey of LLM-based simulation distinguishes agent construction, alignment, and evaluation challenges \citep{gao2024survey}. These works motivate careful agent and runtime descriptions. Their results do not supply external validity for Jevolution's animal-like agents.

\paragraph{Agent-based modeling and testbeds.}
NetLogo and Mesa already provide environments and components for constructing agent-based simulations \citep{wilensky1999netlogo,masad2015mesa}. The ODD protocol and its updates organize model purpose, entities, scheduling, initialization, inputs, and submodels to support understanding and reimplementation \citep{grimm2006odd,grimm2010odd,grimm2020odd}. The flame framework, linked to the AgentTorch implementation, supports differentiable agent-based models for calibration and action learning \citep{chopra2024flame}. Jevolution does not implement that learning method. MAgent and Neural MMO supply many-agent settings, including spatial interaction and competition for resources \citep{zheng2018magent,suarez2019neuralmmo}. Melting Pot emphasizes evaluation with unfamiliar partners and social situations \citep{leibo2021meltingpot}; PettingZoo makes multi-agent interaction order explicit through a standard API and agent-environment-cycle model \citep{terry2021pettingzoo}. We make no claim to exceed these platforms in scale or evaluation breadth.

\paragraph{Communication and adaptation.}
RIAL/DIAL, CommNet, and learned referential games optimize communication along with agent behavior \citep{foerster2016learning,sukhbaatar2016commnet,lazaridou2016language}. Grounded multi-agent communication has also been studied in physical environments \citep{mordatch2018grounded}. Jevolution instead exposes a predefined vocabulary to a remotely served policy. Message use is therefore distinct from discovering a new language. Following \citet{lowe2019pitfalls}, a correlation between a signal and an action is insufficient evidence of receiver influence. In biological communication, playback of different vervet alarm calls elicited different defensive responses without a predator being present \citep{seyfarth1980alarm}; this illustrates an intervention on received information, not an equivalence between text symbols and natural signals. Similarly, hide-and-seek autocurricula train policies \citep{baker2020emergent}, and Evolutionary Policy Optimization combines population search with policy gradients \citep{wang2025epo}. Neither is the same operation as mutating an offspring's simulation traits while leaving policy training outside the experiment.

\paragraph{Ecological context.}
Lotka and Volterra analyze coupled populations and cyclic solutions under explicit assumptions \citep{lotka1925biology,volterra1926fluctuations}. Controlled population experiments by Gause and spatial predator--prey experiments by Huffaker expose the importance of experimental conditions, dispersion, and access to prey \citep{gause1934struggle,huffaker1958predation}. Rosenzweig and MacArthur show how interaction structure affects stability \citep{rosenzweig1963stability}. These sources motivate explicit ecological mechanisms and restrained interpretation. Jevolution's immigration, terrain, capacity limits, and decision scheduling prevent inferring a Lotka--Volterra mechanism from the appearance of its population traces.

\paragraph{Timing and evaluation.}
Real-Time Reinforcement Learning models an environment that evolves during action selection \citep{ramstedt2019realtime}; random-delay formulations separately describe observation and action delays \citep{ramstedt2020delays}. Thus, reporting only provider response time omits other contributors to stale actions. Work on reproducibility and uncertainty in deep-RL evaluation also warns against treating a few stochastic runs as decisive evidence \citep{henderson2018matters,agarwal2021precipice}. We use these as methodological precedents rather than as evidence that this experiment trains a deep-RL policy.

The distinction in this paper is the specified combination of asynchronous controller calls, demographic turnover, engine-defined interaction mechanisms, and request-linked records in one pinned task. Appendix~\ref{app:comparison} compares documented emphases without inferring absent features or measured superiority. The contribution is an executable reference for these experimental concerns; comparative practical advantage remains untested.

\section{Architecture and experimental workflow}\label{sec:architecture}
\paragraph{World, runtime, controller, and record.}
Here, a runtime lineage means a configured controller group, not an inferred genealogical lineage. Individual parent IDs and generations are separate ancestry fields.

\begin{figure}[htbp]
\centering
\includegraphics[width=\linewidth]{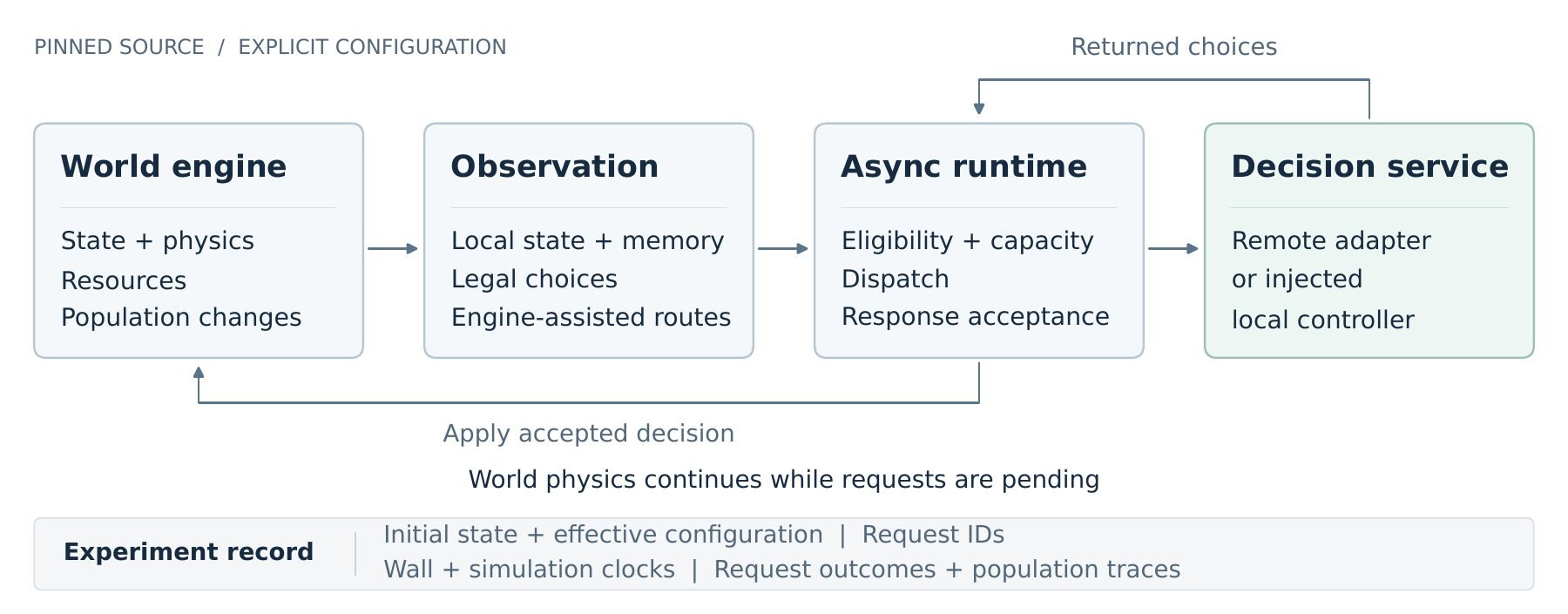}
\caption{Architecture of the pinned implementation. Arrows show information flow, not synchronous execution; world physics continues during requests. The controller receives local observations and engine-offered choices; the engine resolves routes and applies effects. Request identifiers join records across both paths. This schematic is not an empirical result, so no seed count or confidence interval applies.}
\label{fig:architecture}
\end{figure}

The reference architecture has four responsibilities (Figure~\ref{fig:architecture}). The world stores spatial and demographic state and applies physics. The runtime advances that world, identifies eligible actors, and manages outstanding requests. A controller adapter receives a local observation and legal choices and returns a structured decision. Experiment hooks record observations, outcomes, state summaries, and provenance. These boundaries allow a synthetic controller to exercise scheduling without contacting a provider, and allow recorded executions to be analyzed without rerunning inference.

\begin{figure}[htbp]
\centering
\includegraphics[width=\linewidth]{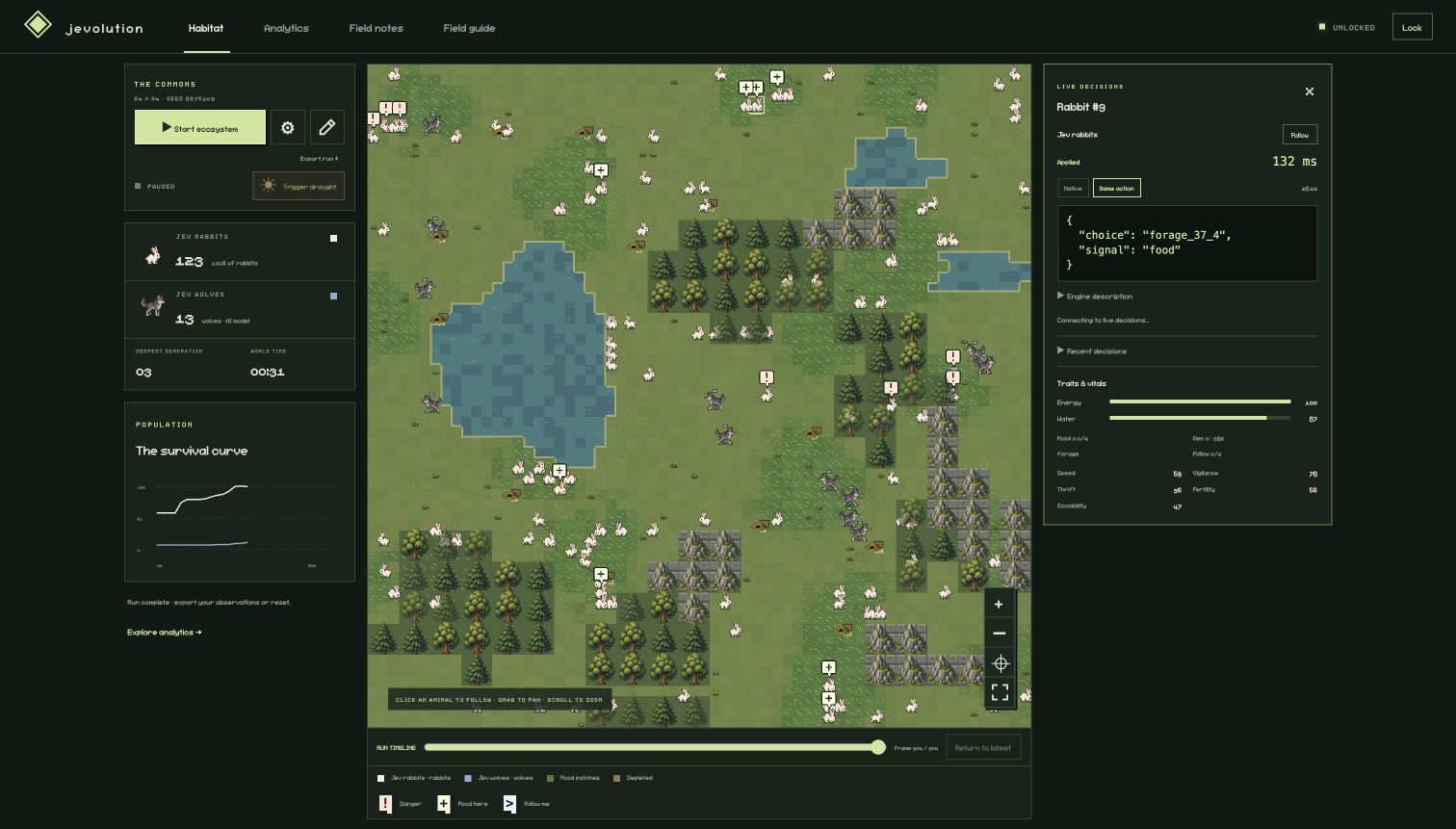}
\caption{Jevolution's interactive interface. Left: run controls and population history. Center: habitat, animals, and signal markers. Right: the selected animal's decision and state inspector. The image is reproduced unchanged from Figure 1 of the authors' original manuscript, at 31.59 simulation seconds; it is not one of the twelve 90-s study runs. Visible signals illustrate the interface, not evidence of a communication benefit. Interface and artwork provenance are documented in Section~\ref{sec:disclosure} and Appendix~\ref{app:artifact}.}
\label{fig:interface}
\end{figure}

\paragraph{Inspecting the world and its decisions.}
The interface brings three views of a run together: population history, the spatial world, and a selected animal's state and decision (Figure~\ref{fig:interface}). This makes it possible to inspect a local decision alongside the world in which it occurs. The screenshot illustrates these views in the historical prototype; the displayed deployment has not been matched to the frozen experimental revision. For the case study, analysis uses the retained experiment records. Appendix~\ref{app:visual-overview} visualizes saved states from one experimental pair.

\paragraph{Two clocks and three delays.}
Simulation time describes the world; wall time controls request acceptance and service waiting. The reference runtime integrates elapsed time using physics substeps no longer than 50 simulation ms. In the live case study, simulation speed is $1\times$, a returned decision must pass a 2,500-wall-ms acceptance threshold, and at most six requests per lineage may be outstanding. A default 100-simulation-ms post-request cooldown has engine-triggered eligibility resets; it is not an unconditional per-animal minimum. An additional global dispatch spacing of 100 wall ms bounds the case-study request rate. Transport timeout, response acceptance, and scheduling eligibility are distinct mechanisms.

Let $t_e$, $t_d$, and $t_a$ denote eligibility, observation/dispatch, and application times on the simulation clock. Queue delay is $t_d-t_e$; applied observation age is $t_a-t_d$. Separately, provider-adapter elapsed time is measured on the wall clock and includes transport and local handling. A long eligibility queue is not necessarily a stale observation because the observation is captured at dispatch. These definitions prevent treating every delay as model compute or every waiting animal as holding an old decision.

\paragraph{Controller boundary.}
The controller receives local state and engine-generated legal choices, not unrestricted control over physics. Route generation may inspect the map beyond the observation; therefore feasible navigation is partly supplied by the engine. In the evaluated adapter, Jev \code{jev-1.13.0} answers independent action and signal \code{Choice} questions sharing one observation \citep{typesafe2026models,typesafe2026primitives}. Neither answer conditions on the other. The code combines both choices into one simulator decision; returned confidence and per-option probabilities are logged but not used for selection or gating. The provider describes this version as a structured decision model without text-generation training \citep{typesafe2026jaggedness}. Its internal architecture and weights were not independently inspected.

\paragraph{From interactive prototype to experiment.}
A reproducible run specifies the source revision, effective rules, initial world, controller identity, clocks, concurrency, dispatch policy, horizon, intervention, and stopping rules. Source-defined controls must be distinguished from interface previews: a visible control alone does not establish a changed mechanism. In the pinned study implementation, communication-off is enforced in the engine. Request, response, and application/error records share identifiers; population counts are retained after each physics substep, with complete initial/final states and approximately one-second world snapshots. The bounded interface journal is not substituted for this append-only experiment record.

The Node host and browser worker both instantiate this core runtime with different persistence and decision-service implementations. This is shared source, not a demonstrated claim of identical wall-clock behavior across hosts. A type-checked local-controller example and the limits of the virtual-clock harness appear in Appendix~\ref{app:extensions}.

These choices support three operations: changing a mechanism with a source-level test, swapping a controller at the decision boundary, and reanalyzing recorded outcomes. Adding a new species, training algorithm, or ecological model still requires implementation work. Configurability is not evidence that every combination is scientifically meaningful or has been tested.

The artifact's reuse contract is a pinned, executable research implementation, not a stable general-purpose plugin API. Configuration, service substitution, and source changes must be distinguished (Table~\ref{tab:configuration}).
\begin{table}[ht]
\centering\small
\caption{Actual intervention routes on the frozen source. This is a source-backed capability map, not a usability experiment.}
\label{tab:configuration}
\begin{tabular}{>{\raggedright\arraybackslash}p{0.22\linewidth}>{\raggedright\arraybackslash}p{0.69\linewidth}}
\toprule
Route & Supported operation and boundary \\
\midrule
Run configuration & Seed, communication condition, timing settings, and audit runtime options. Predator--prey reset installs a fixed Jev roster: supplied provider, population, and group-delay roster edits are replaced. Arena roster configuration is a separate path.\\
Injected service & A host may replace \code{choose} while keeping the engine and observation builder. Synthetic responses test this boundary; virtual clocks additionally require the white-box harness described in Appendix~\ref{app:extensions}.\\
Source changes & New species, a supported clock port, receiver-only/fixed-trait controls, or policy optimization need additional implementation and validation. Frozen UI preview metadata is not an active experimental control.\\
\bottomrule\end{tabular}
\end{table}

\section{Reference model in ODD form}\label{sec:model}
We organize the reference model by the seven ODD elements \citep{grimm2006odd,grimm2010odd,grimm2020odd}; Appendix~\ref{app:odd} expands the description. The implementation is based on commit \code{9285979e95d67d8f11fa4ff6d9b603b156c84380} plus the frozen experimental patch identified in Appendix~\ref{app:protocol}.

\textbf{Purpose and patterns.} The task exposes interactions between local information, asynchronous decisions, and population outcomes. It is not fitted to a real ecosystem. Clustering or oscillating counts are observations to explain, not automatic validation criteria.

\textbf{Entities, state, and scales.} A $64\times64$ habitat contains terrain and food. Animals have identity, position, energy, age, lineage, activity, and scheduling state. Rabbits also have bounded memory and five scalar traits: speed, vigilance, thrift, fertility, and sociability. Signals have a source, position, type, delivery time, and expiry. The reference predator--prey initialization contains 70 rabbits and 8 wolves; effective parameters are retained per run.

\textbf{Process overview.} Physics updates resource availability, movement, metabolism, predation, reproduction, and immigration while requests are outstanding. Substep order is specified in Appendix~\ref{app:mechanisms}. Newborn insertion and the ordering of rabbit and wolf updates can affect trajectories; asynchronous returned choices follow a separate runtime path.

\textbf{Design concepts.} Policies operate with partial observations, predefined signals, and legal actions. Both species reproduce automatically and from a single parent in this preset. The remotely served controller is not trained inside a run. Inherited simulator traits affect mechanics; they are not inherited model weights. Engine rules, resource geometry, and birth placement can produce spatial structure without message use.

\textbf{Initialization and inputs.} Habitat seeds produce recorded initial states. Paired conditions match biological-state hashes; the full configuration hash includes the intervention. Later trajectories can diverge through choices, completion order, and state-dependent random draws. Input data are synthetic world state and controller responses, not measurements from animals. The source is the implementation specification; saved effective configurations prevent an unversioned preset name from standing in for actual settings.

\textbf{Submodels.} For sociability $s$, signal radius is $3+5s$ and immediate energy cost is $0.6(1+s)$. Delivery is scheduled 0.2 simulation seconds after emission and expiry six seconds after emission. The global history retains at most 200 signals; oldest-entry eviction can end availability earlier than the nominal expiry. Terrain, range, and observation timing restrict exposure. The predator--prey vocabulary is danger, food, follow, and silence; help and wolf scent signaling are disabled in this preset. Child rabbit traits follow
\begin{equation}\label{eq:inheritance}
 g_{\rm child}=\operatorname{clip}_{[0.05,0.95]}(g_{\rm parent}+\epsilon),\qquad
 \epsilon\sim\mathcal U[-0.065,0.065),
\end{equation}
with a finite pseudorandom implementation. Food uses capped linear regrowth, not a logistic population equation. Capacity, energy, age, placement, and cooldown constrain births. Immigration attempts add two rabbits and one wolf after at least 20 simulation seconds since the prior eligible check, subject to placement. Between resets and external edits, counts obey
\begin{equation}\label{eq:accounting}
N(t)=N(0)+B(t)+I(t)-D(t).
\end{equation}
Births $B$ and immigrants $I$ must remain distinct: temporary zero abundance need not be permanent extinction, and abundance is not individual fitness.

\section{Research scope and evidence requirements}\label{sec:questions}
The framework supports a research agenda, but each question needs its own identifying intervention. Table~\ref{tab:scope} separates current mechanisms and evidence from unresolved questions. A broad framework contribution does not turn these questions into completed findings.

\begin{table}[t]
\centering\small
\caption{Research scope of the reference system. These are capability/evidence distinctions, not empirical estimates; no seed count or interval applies. The case study and synthetic checks are reported separately.}
\label{tab:scope}
\begin{tabular}{>{\raggedright\arraybackslash}p{0.16\linewidth}>{\raggedright\arraybackslash}p{0.35\linewidth}>{\raggedright\arraybackslash}p{0.40\linewidth}}
\toprule
Topic & Implemented basis & Evidence still needed for a scientific claim \\
\midrule
Inference latency & Asynchronous requests, eligibility, deadlines, explicit clocks & A controlled delay intervention with matched policy, observation, and scheduling; population effects are not established by deadline tests.\\
Communication & Predefined messages; engine-enforced on/off mechanism & Receiver-only and cost-matched interventions to separate information use from energy/option changes.\\
Emergent behavior & Interacting agents, spatial resources, births, immigration & Repeated patterns and spatially matched nulls preserving resource attraction and demographic placement.\\
Natural selection & Inherited traits, ancestry state, and immigration counters; cohort reconstruction required & Cohort/lineage reproductive success and fixed-trait or inheritance controls separating mutation, drift, immigration, and selection.\\
Policy learning & Controller adapter boundary & A training loop, learning objective, and evaluation protocol; EPO is not implemented.\\
\bottomrule
\end{tabular}
\end{table}

\paragraph{Timing as an intervention.}
Injecting a known response delay can verify deadline handling under a controlled clock. To infer the causal effect of latency on collective outcomes, however, an experiment must also specify whether observations are captured before or after the delay, how pending calls consume concurrency, and whether the world advances during waiting. A synthetic deadline fixture answers the software question; it is not a population-effect experiment or a measurement of the provider's inference latency.

\paragraph{Communication and emergence.}
Message production, exposure, and influence are different observations. A predefined warning alphabet does not establish learned language. Receiver interventions are needed to assess influence \citep{lowe2019pitfalls}. Likewise, a cluster near food can be generated by shared attraction, births near parents, or immigration beside conspecifics. A proposed emergence study would compare nearest-neighbor or recruitment statistics against spatially matched nulls retaining those mechanisms, and repeat the comparison over run-level replicates. Such null analyses have not been performed here.

\paragraph{Inheritance, natural selection, and EPO.}
Equation~\ref{eq:inheritance} supplies variation and inheritance at the simulator-trait level. A changing average trait is insufficient to attribute adaptive change when arrivals, mutation, mortality, and reproduction all contribute. Genealogical outcomes and interventions are required. A controller-group label is not an ancestry tree; rabbit single-parent births store the same parent ID twice, so parent counts must be deduplicated when defining ancestry. Evolutionary Policy Optimization instead combines population search with policy gradients \citep{wang2025epo}; Jevolution has no such policy-optimization loop. A future implementation would need explicit policy identities, fitness/evaluation rules, and training data beyond the current controller adapter.

\paragraph{Ecological reference rather than ecological equivalence.}
The classical Lotka--Volterra equations describe aggregate populations under explicit assumptions \citep{lotka1925biology,volterra1926fluctuations}. Jevolution instead applies spatial, resource, birth, immigration, and decision rules. We do not fit Lotka--Volterra coefficients or infer those equations from visually cycling trajectories. Gause, Huffaker, and Rosenzweig--MacArthur motivate attention to experimental conditions and interaction structure \citep{gause1934struggle,huffaker1958predation,rosenzweig1963stability}. Ecological predictive use would require external measurements, calibration, and independent validation.

\section{Validation and worked example}\label{sec:evaluation}
\subsection{Software behavior under controlled inputs}\label{sec:validation}
A separately declared free protocol tests the pinned source with deterministic injected responses and a virtual clock. It holds a small artificial world away from ecological interactions, crosses six response delays with two per-lineage capacities and two seeds, and repeats every condition exactly. This yields 48 executions of 24 conditions, plus 4 executions of pause/removal fixtures. The deadline is deliberately set to 1,000 virtual wall ms; these fixtures are not the live Jev configuration. All matrix executions reach the exact 4.137-s horizon, and pending calls at the horizon remain recorded as cancellations. Observation and terminal records join by request ID.

Responses at the deadline apply, whereas those one millisecond beyond it are late (Table~\ref{tab:synthetic}). At a fixed 100-ms response delay, capacity one yielded 74 applied decisions per execution and capacity three yielded 120: the response time alone did not determine decision opportunity in this fixture. These counts are not a live-service throughput benchmark. The outstanding-call caps hold in the tested matrix; every dispatched request has one terminal outcome. Repeated runs match the defined canonical state/trace hash. This is schedule-conditioned reproducibility under controlled inputs, not bitwise repeatability of a remote service or a proof for untested schedules.

\begin{table}[ht]
\centering\small
\caption{Synthetic runtime fixtures, not provider throughput: request/applied/late/cancelled counts per execution over 4.137 simulation seconds. Each cell was run for two seeds (42, 314159), each repeated twice; all four executions in a cell had the shown counts. There is no sampling-based interval for these deterministic fixtures. Capacity is per lineage, with three animals in each of two lineages. The acceptance deadline is 1,000 virtual wall ms, distinct from the live study's 2,500 wall ms.}
\label{tab:synthetic}
\begin{tabular}{rcc}\toprule
Injected delay (ms) & Capacity 1 & Capacity 3\\\midrule
0 & 238 / 238 / 0 / 0 & 240 / 240 / 0 / 0 \\
100 & 76 / 74 / 0 / 2 & 126 / 120 / 0 / 6 \\
500 & 18 / 16 / 0 / 2 & 42 / 36 / 0 / 6 \\
1000 & 10 / 8 / 0 / 2 & 24 / 18 / 0 / 6 \\
1001 & 10 / 0 / 8 / 2 & 24 / 0 / 18 / 6 \\
1500 & 6 / 0 / 4 / 2 & 18 / 0 / 12 / 6 \\
\bottomrule\end{tabular}
\end{table}

Separate fixtures cover 72 signal boundary cells, 96 reproduction conditions (each repeated), and 16 empty-population immigration conditions (each repeated). They verify the tested delivery/range/expiry rules, mutation bounds and ancestry, and arrival accounting. All 2,724 recorded invariant checks pass in the completed suite. One earlier harness attempt stalled on a terminal floating-point clock boundary; its partial records and the harness correction are retained. No engine source changed. Appendix~\ref{app:validation} gives the fixtures and provenance. These results establish specific software behavior, not adaptation, model competence, external usability, or ecological validity.

\subsection{Communication as an end-to-end case study}\label{sec:case}
The worked example asks whether enabling the implemented rabbit signal system changes population outcomes with Jev controlling both species. It reuses the completed, locally preregistered study; it is not new data collected for this framing. Each of six habitat seeds has one signals-on and one signals-off run, with identical initial biological-state hashes and a fixed 90-s horizon. All twelve main runs reached that horizon; no pair was excluded or replaced. The order was randomized once, yielding five on-first pairs and one off-first pair. No drought was applied. Appendix~\ref{app:protocol} retains all seeds, order, source, and stopping rules.

Signals-off blocks non-silent emission, reception, signal-derived choices, and signal energy charges while preserving personal visual memory and visually supported following. Thus the intervention changes the entire signal system, not solely the information value of messages. The endpoint is rabbit population area under the curve divided by $90\times70$, using left-held post-update counts from every engine substep. It includes descendants and immigrants and is not individual fitness. The estimator is the mean within-seed on-minus-off difference. The nominal interval resamples whole seed pairs 10,000 times with analysis seed 20260928 \citep{efron1979bootstrap}. Animals and requests are not independent replicates.

Across 6 complete habitat-seed pairs, the mean signals-on minus signals-off difference in normalized rabbit abundance was -0.0095 (nominal 95\% pair-bootstrap interval [-0.0772, 0.0602]). This small study does not resolve the direction of the signal-system effect.

\begin{figure}[ht]
\centering
\includegraphics[width=0.88\linewidth]{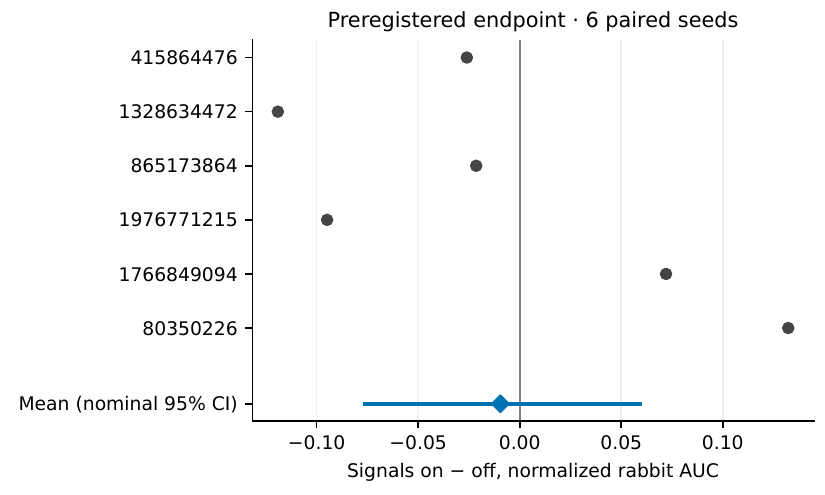}
\caption{Communication worked example: all $n=6$ paired normalized rabbit-AUC effects and their mean. Each pair contains one 90-s run per arm. The nominal 95\% percentile-bootstrap interval uses 10,000 whole-pair resamples (seed 20260928); few-pair coverage is unreliable. These are the same recorded runs as the communication-focused draft, not an independent replication.}
\label{fig:case}
\end{figure}

The leave-one-pair-out means range from -0.0379 to 0.0124. This sensitivity and the interval spanning zero limit the case-study conclusion. The contribution to the framework argument is the traceable route from configuration through requests and mechanism intervention to a qualified run-level estimate, not a positive communication result.

\paragraph{Operational evidence and cost.}
Main runs submitted 10,514 requests, of which 10,462 were applied. The full inventory retains late, invalid, error, and cancelled outcomes, including transport failures, rather than dropping their runs. Across all paid attempts, including pilots and unsuccessful calls, returned usage corresponds to \$1.226087; 24 requests without settled usage retain \$0.066060 in full-context reservations, giving \$1.292148 in conservative task accounting. No additional paid inference was used for the framework validation. These are token-derived charges and bounded liabilities, not an independently reconciled per-run invoice. The shared attempt inventory retains the two pilots, two engineering smoke runs, twelve free-controller runs, and twelve main runs. The free controller ignores received messages and has zero provider latency; it is a separate descriptive reference, not a matched Jev comparison.

The retained timing records distinguish eligibility waiting from observation age and provider-adapter elapsed time (Appendix~\ref{app:timing-exposure}). On-condition rabbit snapshots expose received messages in 828 of 2,634 observations; off-condition snapshots do so in zero of 2,601. These are exposure observations, not distinct deliveries or evidence of listening. The case study establishes neither learned language nor adaptive selection, a causal population effect of latency, or ecological predictive validity.

\section{Limitations and intended use}\label{sec:limits}
The framework has one fully instrumented, remotely controlled case study in one preset. The synthetic fixtures test selected implementation contracts; they do not measure biological validity, service performance, scalability, developer productivity, or usefulness to independent users. Other provider adapters in the repository are implementation seams, not evidence of interchangeable model capability. No controlled platform-to-platform benchmark, load sweep on live services, or evaluation across environmental families is reported.

The communication intervention changes costs, observations, and offered choices together. Six paired habitat seeds and one execution per condition cannot separate service variability from treatment-by-habitat variation. Five pairs happened to run signals-on first; temporal service drift remains a possible confound. Short horizons, engine-assisted navigation, density-dependent births, immigration, and the dispatch throttle constrain interpretation. Numerical-step sensitivity is untested: predation interference is sampled per physics substep without a duration-scaled hazard, so different step partitions can change encounter opportunities. The synthetic timing fixtures suppress predation and cannot bound this pathway. No extra paid condition was added after examining the existing results. The new synthetic validation has its own protocol and does not alter the frozen communication analysis.

Source inspection identifies implemented behavior but does not prove that an old deployment used the same revision. Recorded responses support reanalysis of an execution, not bitwise repetition of a changing remote service. Earlier exploratory exports and free-controller runs have weaker source provenance and remain separate. Logs are not a complete world state at every frame. Public archival release, a full dependency/service-license inventory, sponsor-role confirmation, and independent usability evaluation remain unfinished release or research work.

\section{Broader impact and disclosure}\label{sec:disclosure}
Jevolution can support research on how timing, observations, and interaction mechanisms affect agent evaluation. Its synthetic setting also creates a risk: visually plausible animals and population curves can be mistaken for ecological evidence. All results should be identified as artificial, and ecological applications would require separate calibration and external validation. No live animals or human participants are involved. Any release of prompts, logs, and source must exclude credentials and private information. Bounded request spending and retained failures make the cost of the case study visible; no energy-efficiency claim is made.

\ifdefined\anonymousdraft
\textbf{Disclosure status.} Identifying acknowledgments are withheld in this anonymous working copy and must be restored and confirmed before public release.
\else
\textbf{Author-reported relationships.} The original manuscript states that the prototype received Jevathon prizes sponsored by TypeSafe and Cognition, that CodeRabbit provided development resources, and that all three authors interned at CodeRabbit. Sponsor involvement in study design, analysis, and publication decisions is not documented in the available materials.
\fi

\textbf{Provenance and assistance.} The repository retains the Pixel Agents package identity and an MIT notice crediting Pablo De Lucca.\footnote{\url{https://github.com/pixel-agents-hq/pixel-agents}} Inherited interface infrastructure and assets are not claimed as new contributions. An AI coding assistant supported source inspection, literature verification, experiment tooling, and writing. Authors remain responsible for claims and disclosures. The local artifact contains both this framework framing and an alternative communication-focused draft using the same case-study data; they are not independent studies or two independent empirical contributions. This manuscript is a preprint; no venue acceptance is claimed.

\bibliographystyle{plainnat}
\bibliography{references}
\appendix
\clearpage
\section{Recorded-state illustration}\label{app:visual-overview}
\begin{figure}[htbp]
\centering
\includegraphics[width=\linewidth]{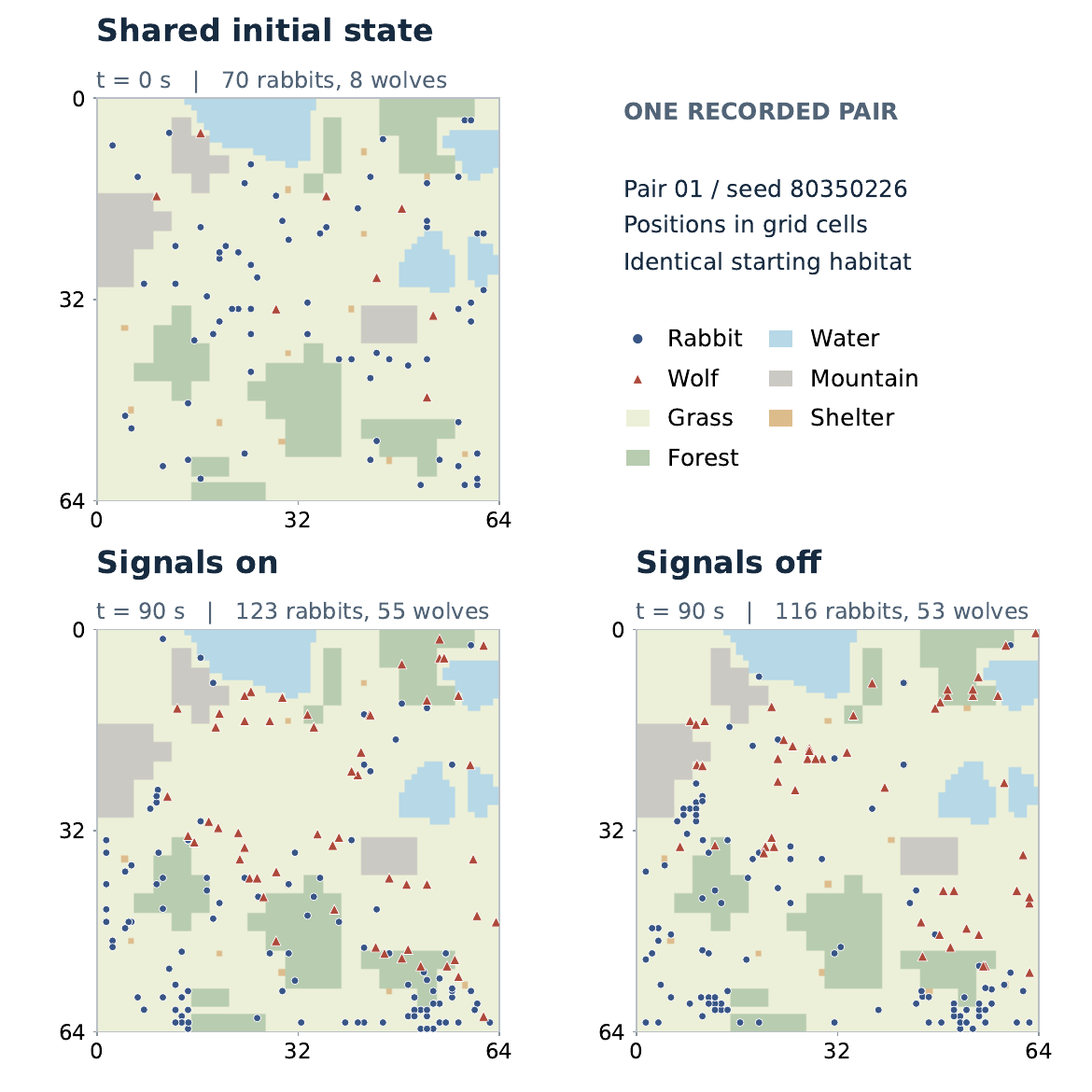}
\caption{Coordinates from retained world states for pair 01 (habitat seed 80350226): the shared initial state and the two terminal states at 90 simulation seconds. This is a visualization of recorded data, not a screenshot captured during execution. It shows one of the six pairs ($n=1$ illustrated pair, two runs); the first numbered pair was chosen by index during figure preparation, after data collection, rather than by its effect or appearance. The terrain and initial animal coordinates match across arms. Markers show positions; overlapping markers can hide individuals. Food, energy, signals, and ancestry are omitted. No spatial hypothesis or clustering statistic is tested, so no confidence interval is attached; the aggregate paired estimate and uncertainty are reported in Figure~\ref{fig:case}.}
\label{fig:recorded-states}
\end{figure}

Figure~\ref{fig:recorded-states} plots coordinates directly from the retained initial and final records of the first numbered experimental pair. Unlike the historical interface illustration in Figure~\ref{fig:interface}, these states come from the paired study. They illustrate the spatial setting; neither a visually coherent trajectory nor a visible signal establishes that a recipient used a message or that communication improved an outcome.

The plot reads \code{initial.json} and \code{final.json} for \code{main-pair-01-signals\_on}, and \code{final.json} for \code{main-pair-01-signals\_off}. The renderer verifies the same terrain and starting animal states in both arms. The retained artifact records the input hashes and panel counts in \code{recorded-state-provenance.json}; \code{render\_recorded\_states.py} regenerates the figure without calling a provider or advancing the simulation.

\clearpage
\section{Position relative to existing systems}\label{app:comparison}
Table~\ref{tab:comparison} lists documented positive properties of neighboring systems. It is a scope comparison, not an exhaustive feature inventory or evidence that an unlisted property is absent. No cross-platform benchmark or ease-of-use experiment was performed.
\begin{table}[ht]
\centering\small
\caption{Documented system emphases and this paper's narrower reference task. These are literature/source descriptions, not measured comparative scores.}
\label{tab:comparison}
\begin{tabular}{>{\raggedright\arraybackslash}p{0.22\linewidth}>{\raggedright\arraybackslash}p{0.32\linewidth}>{\raggedright\arraybackslash}p{0.36\linewidth}}
\toprule
System & Documented emphasis & Relationship to this implementation \\
\midrule
NetLogo / Mesa \citep{wilensky1999netlogo,masad2015mesa} & Environments and components for constructing agent-based models. & Jevolution supplies one specialized population task and a remote-decision execution/recording path; replacement of general ABM tools is not established.\\
Generative Agents, Project Sid, AgentSociety \citep{park2023generative,altera2024sid,piao2025agentsociety} & Agent architectures, real-time social worlds, and simulation/intervention systems. & Real-time agent societies and framework-plus-intervention studies are prior art. This case uses a structured decision interface in a mechanistic population task.\\
flame / AgentTorch \citep{chopra2024flame} & Differentiable agent-based models for calibration and action learning. & Jevolution does not differentiate through its world or optimize its controller in this study.\\
PettingZoo / Melting Pot \citep{terry2021pettingzoo,leibo2021meltingpot} & Explicit multi-agent interaction API; evaluation across social situations and partners. & The present service boundary is source-level and domain-specific; API compatibility and equivalent evaluation breadth are not established.\\
Jevolution & Pinned spatial engine, asynchronous requests, population turnover, signals, inherited traits, and joined records. & Evidence covers the declared software fixtures and one live-model communication intervention. Practical advantage over the systems above remains an open evaluation question.\\
\bottomrule\end{tabular}
\end{table}

\clearpage
\section{Controller extension and implementation boundaries}\label{app:extensions}

\paragraph{Concrete controller interface.}
The frozen \code{RuntimeServices} interface supplies identifiers, provider readiness and defaults, an asynchronous \code{choose} function, initial-state and event recording, and replay construction. Its decision signature is
\begin{center}\small
\code{choose(group, observation, abortSignal, options): Promise<Result>},
\end{center}
where \code{group} is a \code{ModelGroup}, \code{observation} is the rabbit/wolf \code{ModelObservation} union, and \code{options} contains required Boolean \code{relief} and \code{predatorPrey} flags, optional \code{communication} (\code{on}/\code{off}), and optional \code{requestId}. The result carries a legal choice identifier, signal, model identifier, input/output token counts, provider elapsed time, and a nullable provider request identifier. This boundary is defined in \path{core/src/evolution/runtimePorts.ts}, lines 18--33; the observation and result types are in \path{core/src/evolution/types.ts}, lines 179--260.

The following minimal local controller was type-checked against those types. It illustrates interface substitution only; it is not a newly evaluated baseline. The import of \code{RuntimeServices} is omitted for space.
{\small
\begin{verbatim}
const localChoice: RuntimeServices['choose'] =
  async (_group, observation, _abort, _options) => {
    const rest = observation.choices.find(c => c.id === 'rest');
    if (!rest) throw new Error('No rest option');
    return {
      decision: { choice: rest.id, signal: 'none' },
      model: 'offline-rest-v1', inputTokens: 0, outputTokens: 0,
      latencyMs: 0, requestId: null,
    };
  };
\end{verbatim}}
Supply this function as \code{choose} in the service object passed to \code{new SimulationRuntime(services, auditOptions)}. An offline host also supplies readiness for its configured groups, deterministic identifiers, and recording/replay services. It should forbid network access explicitly. The engine and observation builder remain unchanged. A more useful controller must handle species-specific observations, choose only supplied actions, and respect the requested communication condition.

\paragraph{Virtual time is a testing technique.}
The frozen service interface has no clock parameter. Existing offline harnesses replace monotonic-clock and timer globals in a separate process and invoke the private runtime tick through a test-only structural cast; pending promise completions must be scheduled explicitly. The earlier free controller harness demonstrates this at \path{scripts/run_free_controls.ts}, lines 69--82 and 170--216, in the communication artifact. This avoids engine edits but is a white-box harness, not a stable public virtual-clock API. Global functions are restored afterward. A supported clock port would require a future runtime-interface change. Synthetic timing results therefore describe the specified virtual schedule, not real provider latency.

\paragraph{Extensions requiring new implementation.}
A new species requires extending the species/observation/action types, roster validation, state initialization, observation construction, decision application, lifecycle and runtime dispatch, plus corresponding logging, tests, and any desired interface rendering. The present validator allows only rabbits and wolves (\path{constants.ts}, lines 342--395, in the core evolution directory); runtime dispatch distinguishes them by the presence of rabbit genes (\path{runtime.ts}, lines 346--359). Adding a provider instead extends the adapter and configuration paths while preserving the decision contract (\path{server/src/evolution/models.ts}, lines 115--316).

Evolutionary Policy Optimization would require a separate policy representation, evaluation/fitness protocol, policy-variation and selection operators, and persistent generation-level checkpoints and compute accounting. None is supplied by rabbit trait mutation. The current inheritance rule changes five simulator scalars (\path{core/src/evolution/simulation.ts}, lines 435--484), while the controller service remains fixed during a run. Such an optimizer would be a new research implementation requiring its own controls and validation.

\clearpage
\section{Expanded ODD description}\label{app:odd}
We use the seven ODD elements \citep{grimm2020odd}. The experimental implementation is based on commit \code{9285979e95d67d8f11fa4ff6d9b603b156c84380}, with the frozen patch identified in Appendix~\ref{app:protocol}. It corrects reproduction instructions, enforces communication-off at the engine, adds logging and spending controls, and bounds dispatch frequency. Historical exports do not establish that they were generated by precisely this revision and are not pooled with the new study.

\subsection{Purpose and patterns}
The model is an artificial decision-making task for examining how local information and asynchronous execution affect simulated population outcomes. The completed remote-controller evaluation concerns a short-horizon communication intervention; the framework also receives separate synthetic software validation. It does not target any measured rabbit or wolf population, fit ecological parameters, or predict conservation outcomes. Visual clustering, repeated population fluctuations, and trait shifts are not validation criteria without additional controls.

\subsection{Entities, state variables, and scales}
The reference habitat is a $64\times64$ grid with terrain and food resources. Animal state includes identity, position, energy, age, lineage, current activity, and decision scheduling. Rabbits also carry bounded memory and five scalar traits: speed, vigilance, thrift, fertility, and sociability. Signals store their source, location, type, delivery time, and expiry. Terrain is fixed during an unedited run; food and animal states evolve. Simulation time and wall time are retained separately in the logs.

\subsection{Process overview and scheduling}
The engine advances resources, movements, metabolic state, predation, reproduction, and immigration while model requests are outstanding. A request packages the animal's local observation and offered actions; a returned choice is normalized and may be rejected or no longer applicable. In the reference runtime, the $2{,}500$-ms acceptance threshold is measured in wall time, while the default $100$-ms post-request cooldown is measured in simulation time and permits engine-triggered eligibility resets. The reference concurrency limit of six is per lineage, not per animal or globally. The transport timeout is separate from the acceptance threshold. Exact process order, fallback behavior, invalid-choice handling, and any experimental dispatch throttle are part of the configuration, not incidental implementation details.

\subsection{Design concepts}
Agents have partial observations, but engine functions may consult the full map when generating legal routes. This distinction prevents attributing all feasible navigation to the model. Signals use a fixed vocabulary: danger, food, follow, and silence in the predator--prey scenario; the help mechanism is disabled and wolves do not signal. The provider selects structured choices from the offered actions and signals, rather than composing an unrestricted message. A received message does not mechanically force a response. Visually supported following can occur without a message and remains available in the off condition.

The evaluated controller is TypeSafe's Jev \code{jev-1.13.0}, which accepts text or structured text state and returns typed decisions \citep{typesafe2026models}. The provider describes this version as a decision model without text-generation training \citep{typesafe2026jaggedness}. Action and signal use separate \code{Choice} questions over one shared observation. The adapter uses each returned \code{choice}; returned confidence and probability fields are logged but do not gate, weight, or resample decisions. Appendix~\ref{app:controller-interface} gives the request schema and code anchors. The study did not independently inspect model architecture, training data, or weights, and its result concerns this served decision interface.

The controller is not trained within a run. A fixed requested model identifier records the intended remote policy; it cannot independently prove that a vendor's underlying weights remained immutable. Rabbit offspring inherit mutated traits, while immigrants receive new traits. Consequently, the living population's trait distribution can change through reproduction, mutation, mortality, immigration, and stochastic variation. A shift alone does not establish adaptive selection.

\subsection{Initialization}
The predator--prey preset creates 70 rabbits and 8 wolves. Pairing uses identical treatment-invariant initial biological state: habitat, resources, animal IDs, positions and traits, and initial engine random-number state. A separate full-condition hash records the intervention. Equal seeds alone do not guarantee identical later random events after action choices and request completion order diverge. Terrain, initial animal state, and hashes are saved rather than reconstructed from an unversioned preset name.

\subsection{Input data}
The experiment supplies remote model responses rather than observations of real animals. Append-only logs retain requests, returned model identifiers, token usage, errors, cancellations, timestamps, and normalized choices, with shared request IDs. Population counts are logged after each engine substep, which is at most 50 simulation ms; complete initial/final states and approximately one-second world snapshots are also retained. This is not a full-state replay at every frame. Substep endpoint differences can miss an animal born and killed within that substep, although cumulative demographic counters retain those events. API credentials are excluded from the artifact. No empirical ecological calibration data enter this prototype.

\subsection{Submodels}
For sociability $s$, the reference signal radius and immediate energy cost are
\begin{equation}
  r(s)=3+5s, \qquad c(s)=0.6(1+s).
\end{equation}
Delivery is scheduled $0.2$ simulation seconds after emission; expiry is six simulation seconds after emission. The global history retains at most 200 signals; oldest-entry eviction can end availability earlier than the nominal expiry. Thus, absent history eviction, the nominal available interval is $5.8$ seconds, and intermittent observations can shorten actual exposure. Range, terrain visibility, and sender identity further restrict reception. The signal location is its emission location.

In this preset, reproduction is automatic and single-parent for both species. For a rabbit trait $g$, the reference inheritance rule is
\begin{equation}
g_{\mathrm{child}}=\operatorname{clip}_{[0.05,0.95]}(g_{\mathrm{parent}}+\epsilon),
\qquad \epsilon\sim\mathcal{U}[-0.065,0.065),
\end{equation}
where the implementation uses a finite pseudorandom generator. This describes a simulator parameter mutation, not a policy-gradient update. Energy, age, capacity, placement, and dispersal rules constrain births. Rabbit requests describe automatic reproduction, and the experimental patch removes the earlier contradictory mutual-mating option.

Food follows capped linear regrowth with terrain-dependent rules; it is not a logistic population equation. The predator--prey preset uses reference food capacity 8 and normal regrowth coefficient 0.1. No drought is triggered in the study. Immigration attempts add two rabbits and one wolf after at least 20 simulation seconds since the prior eligible check, subject to placement. These arrivals are separate from births and may occur after a species reaches zero. Between resets and external edits, population accounting requires
\begin{equation}
N(t)=N(0)+B(t)+I(t)-D(t),
\end{equation}
with births $B$, immigration $I$, and deaths $D$. A zero count is therefore not necessarily permanent extinction.

Appendix~\ref{app:mechanisms} specifies effective recruitment and predation parameters and the ordered physics substeps, including density-dependent and stationary-only rabbit breeding. These are artificial task settings, not calibrated biological rates.

\clearpage
\section{Synthetic validation design and complete-attempt reporting}\label{app:validation}
These fixtures are separate from the remotely served communication study. They make no network calls and load no credentials. Runtime fixtures use a deterministic \code{rest}/\code{none} controller labeled \code{synthetic-fixture}, with zero provider tokens; signal/reproduction fixtures invoke their named mechanisms directly. Source hashes and a protocol were saved before the execution sequence. The protocol, runner versions, full records, and checks are retained under \path{validation/}.

\paragraph{Clock and runtime matrix.}
The runner crosses seeds 42 and 314159, response delays 0, 100, 500, 1,000, 1,001, and 1,500 ms, and outstanding capacities 1 and 3 per lineage. The fixture uses realtime scheduling, a 1,000-ms acceptance deadline, 100-ms post-request interval, no additional global throttle, and a 4.137-s horizon. One simulation second corresponds to 1,000 virtual wall ms. A 10-ms grid is augmented by response/timer events; physics ticks precede response resolution at the same timestamp. Virtual time uses the white-box harness described in Appendix~\ref{app:extensions}, not a public clock API.

The predator--prey preset is used as a container and then deliberately edited: empty grass terrain, three rabbits and three wolves placed far apart, initial energy/water 90, age zero, no initial paths, and breeding/immigration suppressed. These modified fixtures are not ecological replications. An applied observation's age is checked against the injected delay. Count identities require requests = applied + late + invalid + errors + cancelled; caps, complete joins, terminal settlement, and exact rerun hashes are checked. A pause at virtual wall time 300 ms and removal of a pending rabbit at that time separately exercise cancellation. Pause occurs before the next physics tick, so its recorded simulation horizon is 0.29 s; it is an intentional stop, not a failed 4.137-s run.

\paragraph{Signals.}
At sociability 0.05, 0.50, and 0.95, a stationary sender with ample energy emits \code{follow}, or attempts it with signaling disabled. Receivers are positioned at radius minus 0.001, exactly at radius, and radius plus 0.001, and observed at 0.199, 0.200, 5.999, and 6.000 simulation seconds. All 72 on/off observation cells are retained. Checks cover the implemented cost and radius, inclusive range boundary, delivery eligibility at 0.2 s, and expiry at 6 s. Off-mode emission is rejected without changing energy or signal counters. Terrain occlusion and model listening are not tested.

\paragraph{Inheritance and arrivals.}
Seeds 1--16 are crossed with three parental gene profiles (both 0.05; 0.20 and 0.80; both 0.95) and arena versus predator--prey scenarios. The two mature, well-fed, same-lineage parents have generations 2 and 5. Arena uses reciprocal mate actions; predator--prey retains automatic single-parent reproduction. Wolves, immigration, and unrelated spatial constraints are suppressed before one 50-ms physics step. All 192 executions remain in the inventory. Across the 96 first executions, there are 141 offspring records and 0 executions without a birth; exact repeats are excluded from these two counts. These are fixture counts, not estimated reproductive rates.

For each child, its recorded parent IDs determine the appropriate parent mean; its generation must be one greater than the largest recorded parent generation. Each gene must be in [0.05,0.95] and within 0.065 of that parent mean before clipping. Repeated initial and final states are hash-compared. Checking bounds and repeatability does not test the full mutation distribution or establish fitness differences. A separate repeated empty-population fixture for the same 16 seeds crosses the immigration timer and checks two rabbits plus one wolf, empty parent lists, generation zero, and counters. Generation zero alone does not distinguish a founder from an immigrant; timing and arrival records provide that context.

\paragraph{Attempt history and scope.}
An initial launcher encountered the sandbox's IPC restriction before the runner executed; a Node loader entrypoint avoided that limitation. The first executed harness then stopped at the terminal boundary because floating-point simulation time was just below the horizon while virtual wall time already equaled it. Partial records and the original runner are retained. The correction lets virtual time advance so existing horizon clipping completes; source, factors, and criteria are unchanged. The completed attempt reports 2,724 passing checks and no failing check. A clean full-process rerun matched every runtime summary and all saved signal, inheritance, immigration, and invariant arrays. Separate raw-record arithmetic checks also passed; neither is a new biological sample. No paid API call was made.

\clearpage
\section{Frozen protocol and artifact identifiers}\label{app:protocol}
The main-study record was frozen locally at \code{2026-09-28T23:30:56.062648+00:00}. Its SHA-256 is
\begin{quote}\small\path{d023029c8288e9c7571667ac9b2ee542dc6fe49c5b8189cb1c693c14bce8e791}\end{quote}
The source base is commit \code{9285979e95d67d8f11fa4ff6d9b603b156c84380}. The applied implementation patch has SHA-256
\begin{quote}\small\path{98135f920a582b2b6c3117340cdcceea6cedaffc5b7c9424451fef727c956bfb}\end{quote}
These identities distinguish the experimental implementation from the historical illustrative exports and from the unused newer branch. The local bundle contains the patch, dependency lockfile, effective rules, source/script hashes, and study-design manifest. No independently archived dataset DOI is claimed.

\begin{table}[h]
\centering
\caption{Frozen main-study seed blocks and order. Six pairs were planned, with one attempted run per arm and no outcome-selected replacements. Actual completion is reported in the run ledger.}
\begin{tabular}{rll}\toprule
Pair & Habitat seed & First arm\\\midrule
1 & 80350226 & Signals on\\
2 & 1766849094 & Signals on\\
3 & 1976771215 & Signals off\\
4 & 865173864 & Signals on\\
5 & 1328634472 & Signals on\\
6 & 415864476 & Signals on\\\bottomrule
\end{tabular}
\end{table}

Seeds were sampled without replacement from integers 1 through 2147483647 with Python \code{random.Random(2026092801)}. First-arm bits used a separate generator initialized with 2026092802. The realized five-to-one order imbalance was not rerandomized. The primary analysis generator uses seed 20260928. These identifiers specify the intended randomization rather than guarantee identical asynchronous remote-model outputs.

\paragraph{Stopping and provenance.}
The hard horizon is 90 simulation seconds and the run request ceiling is 902. The global 100-wall-ms dispatch spacing bounds submissions to ten per wall second; the ceiling provides operational headroom and is not an invitation to exceed the horizon. The runtime continues through temporary empty-population states. Requests pending at the horizon are cancelled and their unknown usage is conservatively accounted for. The runtime does not automatically retry a failed transport request. Every later dispatch is separately logged and reserved.

Paired inference requires complete horizons, matching treatment-invariant initial-state hashes, and matching source/configuration/protocol provenance. The integration rule uses every recorded engine substep and never substitutes an observed shorter horizon. The complete-pair estimate is accompanied by exclusions and stop reasons, not silently selected successful trajectories.

\paragraph{Recorded clarification.}
After freezing, source review corrected the protocol's description of the 100-simulation-ms animal delay: it is a post-request cooldown with engine-triggered eligibility resets, including hiding transitions and wolf births. The strict global 100-wall-ms dispatch spacing is unchanged. This was a documentation correction; no executable, intervention, seed, endpoint, or analysis changed, and no interim treatment effects were used. The frozen protocol and its hash remain intact, with the correction recorded separately.


\section{Complete attempt inventory}\label{app:run-inventory}
Every recorded attempt is included below. P denotes paid main pairs, H the handwritten reference, and $+/-$ signals on/off. Pilot and engineering labels follow the analysis inventory order. The machine-readable analysis retains exact run identifiers, input hashes, and stop reasons. The handwritten controller uses immediate decisions and ignores received messages; its outcomes are descriptive, not a latency-matched model comparison. Earlier pilot/free-control patch hashes are retained, but their superseded patch bytes were not independently archived; the final frozen source archive establishes main-study provenance.

\begin{table}[ht]
\centering
\small
\caption{All attempted runs. L/I/E/C denotes late, invalid, error, and cancelled requests. Usage and per-run conservative bounds are derived from the paid-request ledger; free/engineering runs make no paid calls. These are not per-run invoices. Times are observed simulation horizons.}
\label{tab:inventory}
\begin{tabular}{lrrrrrrl}
\toprule
Run & Time (s) & Req. & Applied & L/I/E/C & Usage (\$) & Bound (\$) & Status \\
\midrule
E1+ & 2.0 & 20 & 20 & 0/0/0/0 & 0.0000 & 0.0000 & complete \\
E2- & 2.0 & 20 & 20 & 0/0/0/0 & 0.0000 & 0.0000 & complete \\
H01- & 90.0 & 900 & 900 & 0/0/0/0 & 0.0000 & 0.0000 & complete \\
H01+ & 90.0 & 900 & 900 & 0/0/0/0 & 0.0000 & 0.0000 & complete \\
H02- & 90.0 & 900 & 900 & 0/0/0/0 & 0.0000 & 0.0000 & complete \\
H02+ & 90.0 & 900 & 900 & 0/0/0/0 & 0.0000 & 0.0000 & complete \\
H03- & 90.0 & 900 & 900 & 0/0/0/0 & 0.0000 & 0.0000 & complete \\
H03+ & 90.0 & 900 & 900 & 0/0/0/0 & 0.0000 & 0.0000 & complete \\
H04- & 90.0 & 900 & 900 & 0/0/0/0 & 0.0000 & 0.0000 & complete \\
H04+ & 90.0 & 900 & 900 & 0/0/0/0 & 0.0000 & 0.0000 & complete \\
H05- & 90.0 & 900 & 900 & 0/0/0/0 & 0.0000 & 0.0000 & complete \\
H05+ & 90.0 & 900 & 900 & 0/0/0/0 & 0.0000 & 0.0000 & complete \\
H06- & 90.0 & 900 & 900 & 0/0/0/0 & 0.0000 & 0.0000 & complete \\
H06+ & 90.0 & 900 & 900 & 0/0/0/0 & 0.0000 & 0.0000 & complete \\
P01- & 90.0 & 879 & 876 & 0/1/0/2 & 0.0963 & 0.1018 & complete \\
P01+ & 90.0 & 878 & 874 & 0/3/0/1 & 0.1019 & 0.1046 & complete \\
P02- & 90.0 & 856 & 844 & 0/5/4/3 & 0.0928 & 0.1121 & complete \\
P02+ & 90.0 & 879 & 872 & 0/5/0/2 & 0.1028 & 0.1083 & complete \\
P03- & 90.0 & 881 & 873 & 0/4/0/4 & 0.0967 & 0.1077 & complete \\
P03+ & 90.0 & 877 & 874 & 0/3/0/0 & 0.0971 & 0.0971 & complete \\
P04- & 90.0 & 878 & 876 & 0/1/0/1 & 0.0941 & 0.0968 & complete \\
P04+ & 90.0 & 875 & 874 & 0/0/0/1 & 0.0947 & 0.0975 & complete \\
P05- & 90.0 & 878 & 876 & 0/1/0/1 & 0.0935 & 0.0962 & complete \\
P05+ & 90.0 & 878 & 876 & 0/1/0/1 & 0.0975 & 0.1003 & complete \\
P06- & 90.0 & 878 & 876 & 0/2/0/0 & 0.0979 & 0.0979 & complete \\
P06+ & 90.0 & 877 & 871 & 0/5/0/1 & 0.0996 & 0.1023 & complete \\
Pilot1+ & 30.0 & 290 & 288 & 0/0/0/2 & 0.0293 & 0.0348 & complete \\
Pilot2+ & 30.0 & 294 & 290 & 0/3/0/1 & 0.0319 & 0.0346 & complete \\
\bottomrule
\end{tabular}
\end{table}

\begin{table}[ht]
\centering
\small
\caption{Secondary descriptive outcomes for all 12 main attempts. Wolf AUC is normalized by initial wolf abundance; rabbit founder survival is the fraction of initial rabbit IDs remaining at the fixed horizon. No secondary causal mechanism is identified and no extra hypothesis test is performed.}
\label{tab:secondary}
\begin{tabular}{lrrrrrr}
\toprule
Run & Wolf AUC & Founder surv. & R births & R deaths & R arrivals & Signals \\
\midrule
P01- & 3.3617 & 0.400 & 141 & 103 & 8 & 0 \\
P01+ & 3.1830 & 0.500 & 151 & 106 & 8 & 191 \\
P02- & 3.8189 & 0.386 & 149 & 121 & 8 & 0 \\
P02+ & 3.3332 & 0.457 & 143 & 104 & 8 & 169 \\
P03- & 3.1411 & 0.429 & 147 & 104 & 8 & 0 \\
P03+ & 3.0703 & 0.457 & 135 & 104 & 8 & 222 \\
P04- & 3.6287 & 0.414 & 140 & 120 & 8 & 0 \\
P04+ & 3.7298 & 0.386 & 135 & 113 & 8 & 220 \\
P05- & 3.4950 & 0.514 & 160 & 104 & 8 & 0 \\
P05+ & 3.5623 & 0.429 & 144 & 108 & 8 & 211 \\
P06- & 2.5380 & 0.614 & 129 & 70 & 8 & 0 \\
P06+ & 2.7092 & 0.600 & 131 & 72 & 8 & 155 \\
\bottomrule
\end{tabular}
\end{table}

Across all main runs, cumulative engine counters record 1229 rabbit deaths: 1229 attributed to predation and 0 to starvation. These are simulator event categories, not independently observed biological causes.

\begin{table}[ht]
\centering
\small
\caption{Handwritten reference outcomes for 6 seed blocks. These are descriptive normalized rabbit-AUC values with immediate, zero-provider-latency decisions. The policy ignores received signals; its on/off differences can reflect emission cost. These runs are excluded from the main estimate and are not a matched comparison of model capability.}
\label{tab:free-reference}
\begin{tabular}{lrrr}
\toprule
Pair & On AUC & Off AUC & On $-$ off \\
\midrule
pair-01 & 1.6974 & 1.7245 & -0.0271 \\
pair-02 & 1.6762 & 1.6677 & 0.0085 \\
pair-03 & 1.6028 & 1.5810 & 0.0218 \\
pair-04 & 1.6469 & 1.6192 & 0.0277 \\
pair-05 & 1.6734 & 1.6774 & -0.0039 \\
pair-06 & 1.7273 & 1.7530 & -0.0258 \\
\bottomrule
\end{tabular}
\end{table}

\begin{figure}[p]
\centering
\includegraphics[width=\linewidth]{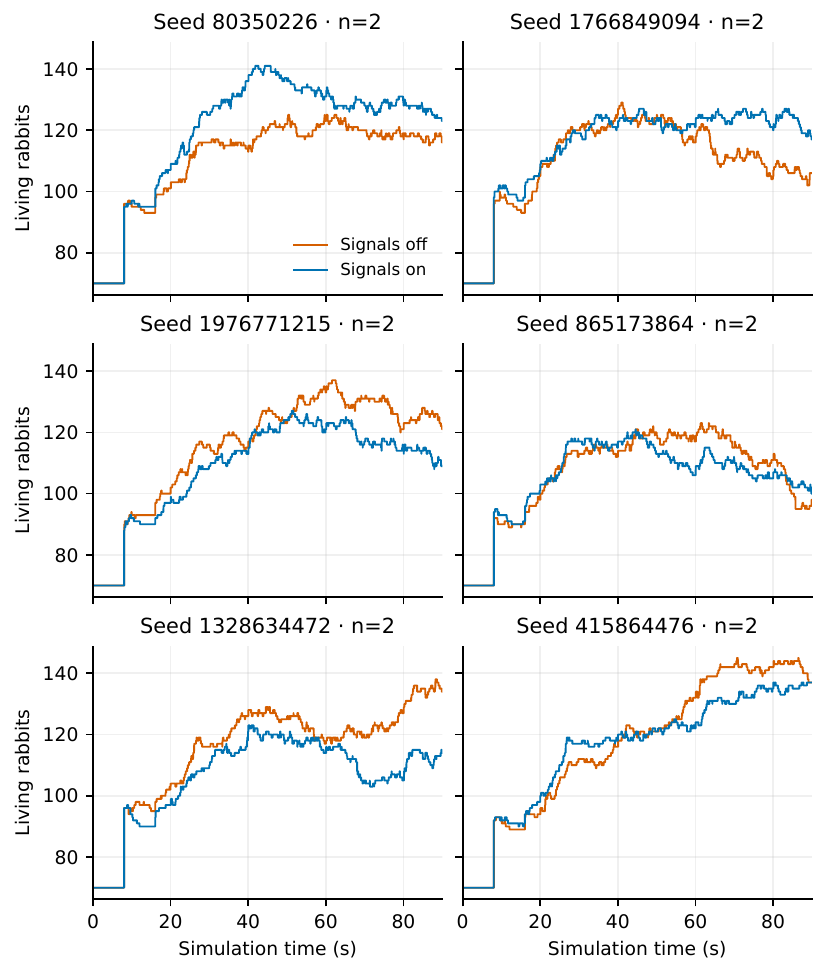}
\caption{All 6 complete main-study seed pairs (12 runs), grouped by the frozen habitat seed. Each panel compares the two recorded arms for one initial state; provider responses and subsequent stochastic paths may differ. These descriptive traces do not establish ecological cycles or message-use mechanisms.}
\label{fig:individual}
\end{figure}
\clearpage
\section{Evaluated controller interface}\label{app:controller-interface}
The completed study uses TypeSafe's \code{jev-1.13.0} through its \code{/v1/systemone} endpoint. The provider documents text inputs, typed decisions, and independently evaluated questions sharing one state \citep{typesafe2026models,typesafe2026primitives}. Its version-specific documentation states that Jev is not trained for text generation \citep{typesafe2026jaggedness}. These are provider descriptions of the served interface. The artifact does not supply model weights, a training corpus, or an independently verified architecture. The experiment therefore characterizes the recorded controller and runtime; it does not evaluate a generative chat-agent architecture.

\paragraph{Request structure.}
The following schematic shows the two questions for a signals-on rabbit. Placeholder state fields, action identifiers, instructions, and abbreviated signal descriptions illustrate the structure; this is not an archived request or the exact experimental prompt. Full requests, including actual legal options and instructions, are retained in each run's \code{requests.jsonl}.
\begin{quote}\small
\begin{minipage}{\linewidth}
\begin{verbatim}
{
  "model": "jev-1.13.0",
  "state": {"...": "local observation fields"},
  "questions": {
    "action": {
      "type": "choice",
      "instructions": "<scenario and condition instructions>",
      "criteria": {"<action ID>": "<legal action description>"}
    },
    "signal": {
      "type": "choice",
      "instructions": "<same instructions; select a signal>",
      "criteria": {
        "none": "<silence criterion>",
        "danger": "<warning criterion>",
        "food": "<food criterion>",
        "follow": "<following criterion>"
      }
    }
  }
}
\end{verbatim}
\end{minipage}
\end{quote}
Both questions see the same observation. Under the documented interface, neither question sees the other's answer. In particular, the signal choice is not conditioned on the action selected in that request. The code combines the two returned choices into one decision for the engine; no second inference call coordinates them. Wolves and communication-off rabbits receive only the \code{none} signal option. This batching preserves one request/application opportunity, but does not establish coordinated optimization over action--signal pairs.

\paragraph{Response handling and source anchors.}
The frozen \code{server/src/evolution/models.ts}, lines 165--179, constructs the two questions; lines 278--298 extract \code{answers.action.choice} and \code{answers.signal.choice}. Lines 99--113 validate these values against the offered action and signal sets. The returned confidence and per-option probabilities are retained in provider-response logs, but the adapter does not threshold, weight, or sample from them. The runtime then checks cancellation, lateness, and current applicability before applying the combined decision (\code{core/src/evolution/runtime.ts}, lines 421--445). These checks concern request timing and engine validity; they are not confidence gates. This description clarifies the frozen implementation and does not modify the policy used to collect the study.

\section{Population submodels and effective task parameters}\label{app:mechanisms}
This supplement expands the ODD summary for the frozen main-study implementation. Values in Table~\ref{tab:mechanisms} are arbitrary simulator task settings, not estimates calibrated to animal populations; no biological or optimality rationale is inferred from them. All 12 main-run metadata records contain the same effective rules. Time below is simulation time, distance is in grid tiles, and energy, water, and food use simulator units. This is a readable specification of the mechanisms most directly affecting the abundance endpoint, not a replacement for the complete frozen code, terrain-generation procedure, or legal-action specification. The companion \code{MECHANISM-APPENDIX-EVIDENCE.md} maps every rule below to source lines and retained configuration.

\begin{table}[ht]
\centering
\small
\caption{Effective parameters governing movement and demography. Trait symbols $g_s,g_v,g_t,f$ denote speed, vigilance, thrift, and fertility. Population caps constrain births, not immigration.}
\label{tab:mechanisms}
\begin{tabular}{@{}p{0.25\linewidth}p{0.70\linewidth}@{}}
\toprule
Mechanism & Main-study setting \\
\midrule
Rabbit recruitment & Maturity 8 s; minimum energy and water 30 each; global crowding scale 160 rabbits; hard birth cap 250; local crowding disabled. \\
Rabbit birth cost & Parent loses $10+12f$ energy; birth or crowding rejection resets cooldown to $16(1.3-0.6f)$ s. Newborn energy/water: 60/80. \\
Wolf recruitment & Maturity 5 s; energy threshold 80; parent cost 30; birth cooldown 10 s; initial founder birth delay 20 s; birth cap 150. Newborn energy: 33. \\
Wolf energy & Metabolism 1.2 energy/s; a kill adds 40, capped at 100; movement energy charge is zero. \\
Movement & Rabbit speed $1.05+0.65g_s-0.16g_t$ tiles/s; wolf speed 1.6 tiles/s. Forest multiplies speed by 0.77; a wolf's eating cooldown adds a factor 0.35. \\
Detection and capture & Rabbit sight radius $4+\lfloor6g_v\rfloor$; wolf sight radius 9, or 2 for a rabbit hiding in forest. Capture distance $<0.8$; eating cooldown 10 s; interference coefficient 1. \\
Newborn dispersal & Up to 12 route attempts; proposed radial distance in $[6,12)$ before boundary clipping and tile centering. Actual displacement and route length need not lie in that interval. \\
Resources & Reference food capacity 8; nondrought regrowth coefficient 0.1; food conversion 5.5 energy/unit; rabbit thirst 0.1 water/s. \\
Lifespan and arrivals & Both lifespan thresholds $10^9$ s. Every eligible 20-s immigration check attempts two rabbits and one wolf, subject to placement. \\
\bottomrule
\end{tabular}
\end{table}

\paragraph{Ordered engine substep.}
The runtime splits elapsed simulation time into substeps of at most 0.05 s, clipping the last substep at the study horizon. Returned model decisions are handled asynchronously; the following is the order within each synchronous physics substep, not an assumption that all animals decide together.
\begin{enumerate}
\item Advance the simulation clock, regrow food, and expire signals (or clear them when signaling is off). On a food-bearing tile of capacity $K$, nondrought regrowth adds $0.1(K/8)\,dt$, capped at $K$.
\item Copy the current rabbit list and rotate its order by a pseudorandom offset. For each rabbit in that order, advance age and cooldown; debit basal energy and water; update following and hiding; move and charge movement energy. If its path is empty, process stationary activities, including foraging, drinking, and automatic breeding. Then remove that rabbit if energy or water is exhausted, or age exceeds its threshold.
\item Process a copy of the current wolf list in its stored order. Advance age and debit metabolism, update cooldowns and pursuit, move, check for a capture, credit a meal if caught, and then remove wolves with exhausted energy or excessive age.
\item Attempt wolf reproduction, then immigration; finish any drought transition and update the bounded display history. The experiment records population counts after the completed substep.
\end{enumerate}
Rabbit newborns are appended during step 2 and are not themselves updated in that copied rabbit loop; they can nevertheless be encountered by wolves in step 3. Wolf newborns enter after the wolf loop. Consequently, births, resource use, and predation are sequential rather than simultaneous events.

\paragraph{Rabbit recruitment and energy.}
A rabbit attempts automatic single-parent reproduction only when its path is empty, age is at least 8 s, cooldown is nonpositive, energy and water are each at least 30, and no wolf is within four tiles (this last test uses distance, not visibility). Let $R$ be the current rabbit count at the attempt. The hard cap first rejects attempts with $R\geq250$. Otherwise a finite-generator draw $u\in[0,1)$ rejects the attempt when $u<R/160$; because the draw lies in $[0,1)$, $R\ge160$ deterministically rejects a birth attempt even below the hard cap of 250. Such a rejection resets the parent's fertility-dependent cooldown without charging birth energy. This is an eligibility-and-cooldown rule, not a continuous logistic population equation. A surviving attempt requires an adjacent walkable location unoccupied by another rabbit; failed placement does not apply the birth cost or cooldown. A successful birth applies the cost and cooldown in the table and clears the parent's current movement/action to rest.

Rabbit basal energy loss per second is $0.1+0.06g_v+0.05g_s-0.035g_t$, multiplied by 0.7 while resting. Movement costs $0.1+0.14g_s$ energy per tile. Stationary foraging consumes up to $1.4(1-0.3g_t)$ food units/s, limited by local food and the energy ceiling of 100; stationary drinking beside water restores up to 16 water units/s, capped at 100. Thus an action can affect abundance through reproduction eligibility and energy as well as predation. This is a possible implemented pathway, not a mechanism identified by the treatment estimate.

\paragraph{Refuges, pursuit, and wolf recruitment.}
Shelter tiles exclude rabbit detection by wolves and exclude capture. Forest hiding reduces detection range but is not absolute protection from an encounter. Visibility additionally tests terrain obstruction. Model-controlled wolves replan an active hunt at intervals of at least 0.65 s while the selected prey remains visible. A wolf can also catch another exposed rabbit it reaches. For a candidate capture with $n$ other wolves within four tiles, the interference check passes when $n=0$ or a pseudorandom draw is below $1/(1+n)$. This check is made within substeps; it is not a fitted continuous-time predation rate. Capture additionally requires a nonpositive eating cooldown, then resets it to 10 s.

Wolf births are automatic and single-parent, requiring the age, energy, cooldown, and cap conditions in the table plus adjacent non-shelter placement unoccupied by any animal. Unlike rabbit births, they do not require the parent to stop moving. Founder wolves begin mature, with energy 50, an eating cooldown of 8 s, and a 20-s birth cooldown. Rabbit founders begin mature, with energy/water 75/90 and an 8-s birth cooldown. Newborns of both species receive dispersal routes if an attempted destination is legal and reachable. Dispersal sets the next decision time to current time plus straight-line destination distance divided by 1.2; this is an eligibility estimate, not a guarantee that the animal has arrived.

\paragraph{Open-population boundary.}
Immigrants are placed beside conspecifics when possible, with an edge-tile fallback. They bypass the birth caps. Arriving rabbits receive fresh trait draws, mature age, and founder resource values; arriving wolves receive founder state except that their birth cooldown is reset to zero. The high lifespan thresholds disable old-age removal over this 90-s horizon. These choices make abundance an outcome of an explicitly open artificial population; they do not establish long-run coexistence, fitness, or a reduction to a classical predator--prey equation.

\clearpage

\section{Post-hoc timing and signal-exposure diagnostics}\label{app:timing-exposure}

These descriptive diagnostics were added after collection in response to reporting review. They use all 12 complete main runs and change no frozen endpoint, inclusion rule, or uncertainty estimate. Rabbit observations contained at least one signal in 828/2634 on-condition requests (31.44\%) and 0/2601 off-condition requests. Exposure means a signal was present in the supplied snapshot; it does not establish attention, listening, behavioral use, or unique delivery events.

Queue delay runs from engine eligibility to dispatch in simulation seconds. Observation age instead runs from the captured observation to application, also in simulation seconds and only for applied decisions. Recorded response time is provider-adapter elapsed wall time for available returns, including transport and local handling; it is not isolated model compute. Missing latency is not imputed as zero. Medians and 95th percentiles use linear interpolation at $(N-1)p$.

Decision-opportunity rates divide requested observations by total living-species animal-minutes, integrated with the frozen left-hold rule and including infants and other animals not yet eligible to act. Thus these are population exposure-normalized rates, not each eligible adult's decision frequency. Decisions and time samples are not independent replicates; no confidence interval, effect test, or causal latency comparison is added.

\begin{table}[p]
\centering
\small
\caption{Signal exposure and population-normalized opportunities. $+/-$ denotes signals on/off. Type columns count rabbit observations containing each kind and are nonexclusive; help is absent. The final columns are rabbit/wolf requests per living animal-minute, not per wall minute or per eligible adult.}
\label{tab:exposure}
\resizebox{\linewidth}{!}{%
\begin{tabular}{lrrrrrrrr}
\toprule
Run & Obs. & Any & \% & Danger & Food & Follow & R req/min & W req/min \\
\midrule
P01- & 440 & 0 & 0.0 & 0 & 0 & 0 & 2.65 & 10.88 \\
P01+ & 439 & 148 & 33.7 & 105 & 46 & 0 & 2.44 & 11.49 \\
P02- & 403 & 0 & 0.0 & 0 & 0 & 0 & 2.46 & 9.88 \\
P02+ & 440 & 121 & 27.5 & 97 & 25 & 0 & 2.56 & 10.98 \\
P03- & 441 & 0 & 0.0 & 0 & 0 & 0 & 2.56 & 11.67 \\
P03+ & 439 & 162 & 36.9 & 146 & 18 & 2 & 2.70 & 11.89 \\
P04- & 439 & 0 & 0.0 & 0 & 0 & 0 & 2.75 & 10.08 \\
P04+ & 438 & 133 & 30.4 & 116 & 15 & 6 & 2.78 & 9.76 \\
P05- & 439 & 0 & 0.0 & 0 & 0 & 0 & 2.56 & 10.47 \\
P05+ & 439 & 149 & 33.9 & 116 & 41 & 0 & 2.76 & 10.27 \\
P06- & 439 & 0 & 0.0 & 0 & 0 & 0 & 2.50 & 14.41 \\
P06+ & 439 & 115 & 26.2 & 105 & 11 & 0 & 2.54 & 13.47 \\
\bottomrule
\end{tabular}}
\end{table}

\begin{table}[p]
\centering
\small
\caption{Per-run/species timing: R denotes rabbits and W wolves. Q is eligibility-to-dispatch queue delay; W in the time header is provider-adapter elapsed wall time. Time entries give median/95th percentile. Queue values are available for every listed observation. Return n/N explicitly excludes missing latency from quantiles; Age n/applied gives the matched-age denominator among applied requests. These are descriptive conditional distributions, not independent experimental samples.}
\label{tab:timing}
\resizebox{\linewidth}{!}{%
\begin{tabular}{lrrrrrr}
\toprule
Run/species & Obs. & Q (sim s) & Return n/N & W (wall ms) & Age n/applied & Age (sim s) \\
\midrule
P01- R & 440 & 18.24/19.49 & 438/440 & 112.5/172.5 & 438/438 & 0.102/0.153 \\
P01- W & 439 & 3.46/7.80 & 439/439 & 107.5/154.9 & 438/438 & 0.101/0.152 \\
P01+ R & 439 & 19.03/22.98 & 439/439 & 116.0/170.9 & 439/439 & 0.102/0.153 \\
P01+ W & 439 & 3.11/7.63 & 438/439 & 107.3/152.1 & 435/435 & 0.101/0.152 \\
P02- R & 403 & 17.36/24.82 & 398/403 & 112.9/161.8 & 398/398 & 0.102/0.153 \\
P02- W & 453 & 3.64/9.44 & 451/453 & 108.6/156.8 & 446/446 & 0.101/0.152 \\
P02+ R & 440 & 18.38/20.49 & 438/440 & 114.8/162.0 & 438/438 & 0.102/0.153 \\
P02+ W & 439 & 3.32/7.75 & 439/439 & 109.1/155.8 & 434/434 & 0.101/0.152 \\
P03- R & 441 & 18.11/21.39 & 437/441 & 116.4/164.7 & 437/437 & 0.102/0.153 \\
P03- W & 440 & 3.47/7.14 & 440/440 & 112.5/165.0 & 436/436 & 0.101/0.153 \\
P03+ R & 439 & 17.39/20.01 & 439/439 & 116.9/165.8 & 438/438 & 0.102/0.153 \\
P03+ W & 438 & 3.51/7.50 & 438/438 & 112.4/161.5 & 436/436 & 0.102/0.153 \\
P04- R & 439 & 17.40/18.85 & 439/439 & 115.7/157.1 & 439/439 & 0.102/0.152 \\
P04- W & 439 & 3.91/9.10 & 438/439 & 112.8/160.7 & 437/437 & 0.101/0.152 \\
P04+ R & 438 & 17.14/19.21 & 437/438 & 117.6/161.0 & 437/437 & 0.102/0.153 \\
P04+ W & 437 & 3.78/8.87 & 437/437 & 111.5/160.5 & 437/437 & 0.101/0.152 \\
P05- R & 439 & 18.02/20.13 & 439/439 & 118.3/183.9 & 439/439 & 0.102/0.153 \\
P05- W & 439 & 4.02/8.73 & 438/439 & 116.8/169.5 & 437/437 & 0.102/0.153 \\
P05+ R & 439 & 17.56/19.02 & 439/439 & 119.5/169.0 & 439/439 & 0.102/0.153 \\
P05+ W & 439 & 3.50/9.09 & 438/439 & 115.4/160.9 & 437/437 & 0.101/0.152 \\
P06- R & 439 & 17.79/22.86 & 439/439 & 121.2/167.1 & 439/439 & 0.102/0.153 \\
P06- W & 439 & 3.20/5.52 & 439/439 & 118.7/158.6 & 437/437 & 0.102/0.152 \\
P06+ R & 439 & 18.39/21.71 & 438/439 & 117.7/159.5 & 438/438 & 0.102/0.152 \\
P06+ W & 438 & 3.26/6.33 & 438/438 & 115.5/152.1 & 433/433 & 0.101/0.105 \\
\bottomrule
\end{tabular}}
\end{table}

\clearpage
\section{Artifact boundaries and reuse}\label{app:artifact}
This manuscript is a framework-oriented alternative to a communication-focused draft prepared from the same Jevolution audit. The communication logs, frozen design, primary estimator, and reported effect are shared evidence. They are not new replications, a second independent study, or grounds for counting the same empirical contribution twice. The new evidence consists of separately declared free synthetic engineering validation and the source-backed architectural specification.

The authors retain a local research package containing the original communication artifact, the separate framework source, the new validation protocol/scripts/results, claim and citation ledgers, and a comparison of the two framings. This package is distinct from the manuscript's typesetting source and is not publicly archived at this revision. The original communication manuscript and frozen inputs remain unchanged. One command regenerates framework figures and compiled manuscripts from these retained inputs; a separately documented free command reruns the engineering suite. Neither invokes remote inference. These are local reproduction checks, not a claim that readers currently have access to the complete package.

The source-backed description covers the pinned experimental implementation. Historical UI exports and the live website are not substituted for that provenance. Bounded historical journals cannot establish complete decision frequencies. The communication run inventory includes all earlier attempts; the synthetic fixtures have a separate attempt inventory and are never pooled into its sample size or uncertainty calculation.

\paragraph{Project links and experimental version.}
\ifdefined\anonymousdraft
Identifying project links are withheld in this anonymous manuscript. The companion source/data package requires separate anonymization before submission.
\else
The project website is \url{https://jevolution.world}, and the public source repository is \url{https://github.com/utkarshg20-org/jevolution}. The recorded base revision is \href{https://github.com/utkarshg20-org/jevolution/commit/9285979e95d67d8f11fa4ff6d9b603b156c84380}{\code{9285979e95d67d8f11fa4ff6d9b603b156c84380}}. A separate retained patch and frozen source archive identify the evaluated implementation (Appendix~\ref{app:protocol}). These links identify the project and base source; they do not substitute for the complete experimental supplement or certify the revision running on the website.
\fi

\paragraph{Documentation accompanying the new assets.}
The retained package's \code{ASSET-CARD.md} documents the code, synthetic records, shared communication logs, schemas, provenance, intended uses, limitations, and license status. Its \code{README.md} gives the exact analysis/build and optional fixture commands. Raw records retain failures and distinguish simulation time from wall time; runtime fixtures are software checks, not independent ecological samples. No new model weights or human-subject dataset is introduced. The arXiv typesetting package contains the manuscript and its figures; it does not distribute the frozen engine, raw service responses, or full experiment dataset. A public repository link alone therefore does not provide independent reproduction of this exact study.

\paragraph{Asset provenance and license scope.}
The frozen source retains the Pixel Agents MIT notice, copyright 2026 Pablo De Lucca. Project records identify the animal and terrain atlases as AI-generated artwork created on 24 September 2026; inherited interface infrastructure and generated artwork are distinct assets. The interface uses FS Pixel Sans Unicode by NZWStudios2024 under SIL Open Font License 1.1.\footnote{Font and license: \url{https://fontstruct.com/fontstructions/show/2606508/fs-pixel-sans-unicode-regular}.} The historical screenshot in Figure~\ref{fig:interface} illustrates the authors' original manuscript; it does not establish the deployed code or artwork revision. The architecture, statistical plots, and recorded-state map are generated by the retained scripts.

Jev is a hosted service subject to provider terms. The currently posted TypeSafe agreement discusses output ownership and also limits specified downstream uses; the account-specific terms governing the experiments have not been verified.\footnote{Current terms inspected 1 October 2026: \url{https://typesafe.ai/legal/mca}.} No unrestricted license for the raw-response dataset is asserted. The MIT source notice, interface-font license, and manuscript-distribution license must not be interpreted as a blanket license for every dependency, artwork, or service output. A complete redistribution review remains separate from reporting aggregate research results.

The full research package remains local. A future release should attach a persistent archive identifier and finalize asset/service and data licensing. Sponsor roles and a future venue's requirements must also be confirmed. An anonymous PDF is not an anonymized raw-data bundle. No public dataset DOI, independent reuse, or venue acceptance is asserted.

\clearpage
\section{Execution resources and accounting}\label{app:compute}
The recorded host was an Apple M5 Pro with 15 logical CPUs, 24 GiB installed memory (25,769,803,776 bytes), macOS (Darwin 25.6.0), and arm64 architecture. Software versions were Node 26.8.1, TypeScript 5.9.3, and tsx 4.23.15. Synthetic manifests independently confirm Node/platform/architecture, without a separate hardware fingerprint. These are host resources, not peak process requirements. Local Node runs the simulator; TypeSafe's remote \code{jev-1.13.0} API supplies paid decisions. No policy training occurs. Paid runs execute serially, with at most six pending requests per lineage and at least 100 wall-ms between shared dispatches.

Table~\ref{tab:compute-elapsed} reports real UTC end-minus-start intervals for all 28 original attempts: 1,153.393 s (19.223 min) total, comprising 1,081.125 s for 12 main runs, 60.246 s for two pilots, 7.900 s for 12 offline handwritten controls, and 4.122 s for two engineering checks. These are elapsed intervals, not CPU-core-seconds, exclusive kernel time, or whole-project duration. The controls' virtual clock advances 90 simulation seconds faster than real time. The paid runs made 11,098 API requests; usage-based cost is \$1.226087394 and conservative accounting including unknown usage is \$1.292147682, not an independently verified invoice.

\begin{table}[ht]
\centering\small
\caption{Recorded elapsed seconds for every original attempt. P: paid main; H: offline handwritten control; E: engineering check. On/off denote the signal mechanism. The horizon column is simulation time; the other columns are real UTC intervals. A dash means no such arm was attempted under that label.}
\label{tab:compute-elapsed}
\begin{tabular}{lrrr}
\toprule
Run & Horizon (sim s) & On (elapsed s) & Off (elapsed s)\\
\midrule
P01 & 90.0 & 90.067 & 90.106 \\
P02 & 90.0 & 90.110 & 90.064 \\
P03 & 90.0 & 90.107 & 90.056 \\
P04 & 90.0 & 90.091 & 90.106 \\
P05 & 90.0 & 90.183 & 90.122 \\
P06 & 90.0 & 90.051 & 90.062 \\
H01 & 90.0 & 0.663 & 0.521 \\
H02 & 90.0 & 0.478 & 0.477 \\
H03 & 90.0 & 1.179 & 0.678 \\
H04 & 90.0 & 1.213 & 0.572 \\
H05 & 90.0 & 0.542 & 0.539 \\
H06 & 90.0 & 0.511 & 0.527 \\
Pilot1 & 30.0 & 30.172 & --- \\
Pilot2 & 30.0 & 30.074 & --- \\
E1 & 2.0 & 2.062 & --- \\
E2 & 2.0 & --- & 2.060 \\
\bottomrule
\end{tabular}
\end{table}

The retained synthetic suite and clean reproduction took 3.868 and 3.173 s between recorded UTC captures. Each includes 52 runtime executions, 72 signal cells, 192 inheritance executions, and 32 immigration executions. Native \code{Date} captures remain real despite virtual \code{Date.now}/\code{performance.now} overrides; they exclude earlier startup and later cleanup. Individual fixture elapsed times were unrecorded. Runtime horizons of 4.137 simulation seconds (pause: 0.290 s) are not elapsed-runtime benchmarks. The launcher failure and incomplete harness execution (Appendix~\ref{app:validation}) have unknown elapsed times, not zero cost.

Original run directories occupy 894,474,997 logical bytes; complete and failed fixture directories occupy 141,942,445 and 22,891 bytes: 1,036,440,333 bytes (1.036 decimal GB) total. Dependencies, extra analysis/build files, and temporary space are additional. The frozen source archive adds 3,279,256 compressed bytes. These are file lengths, not peak disk requirements. Per-run intervals, size definitions, and source-record hashes accompany \path{evidence/compute-resources.json}.

Provider hardware and energy are undisclosed; API access does not require their disclosure. Offline reanalysis uses saved responses; exact future API behavior is not guaranteed. CPU/GPU utilization, peak resident memory/storage, and aggregate development, earlier exploratory work, testing, typesetting, and assistant-service compute were unmeasured. No complete project-wide compute total or hardware-efficiency result is claimed.

\end{document}